# The environmental factors influencing individual decision-making behavior in software projects: a systematic literature review

| Jingdong Jia | Pengnan Zhang | Luiz Fernando Capretz |
|---|---|---|
| School of Software | School of Software | Department of Electrical & Computer |
| Beihang University | Beihang University | Engineering, Western University |
| Beijing, 100191,China | Beijing, 100191,China | London, N6A5B9, Ontario, Canada |
| jiajingdong@buaa.edu.cn | 1206834470@qq.com | lcapretz@uwo.ca |

## ABSTRACT
As one of the crucial human aspects, individual decision-making behavior that may affect the quality of a software project is adaptive to the environment in which the individual is. However, no comprehensive reference framework of the environmental factors influencing individual decision-making behavior in software projects is presently available. This paper undertakes a systematic literature review (SLR) to gain insight into existing studies on this topic. After a careful SLR process, 40 studies were targeted to solve this question. Based on these extracted studies, we first provided a taxonomy of environmental factors comprising eight categories. Then a total of 237 factors are identified and classified using these eight categories, and some major environmental factors of each category are listed in the paper. The environmental factors listing and the taxonomy can help researchers and practitioners to better understand and predict the behavior of individuals during decision making and to design more effective solutions to improve people management in software projects.

## CCS Concepts
• **Social and professional topics~Project and people management**   • **Software and its engineering~Software development process management**   • Software and its engineering~Software configuration management and version control systems   • Applied computing~Law, social and behavioral sciences

## 1. INTRODUCTION
The software development process is a human-centered activity. This fact highlights the effect of human factors in software engineering (SE) [42]. The human factor is a make-or-break issue that affects most software projects [13]. Therefore, it is not surprising to see that the research focusing on human factors in SE has received significant attention. The term "human factor" indicates different aspects of human involvement in software projects [42]. From the personnel structure perspective, the research falls into two categories: the team and the individual. The individual aspect in SE is our focus in this paper because individuals are the primary constituent elements of a team. Regarding this field, many efforts have focused on the individual performance [37], personality [14; 27], abilities and skills [2; 33], attitude [18], and motivation [11; 51]. However, there is a dearth of research focused on individual decision-making behavior. Recently Lenberg et al. has proposed a concept of "behavioral software engineering" by taking cues from behavioral economics [36]. Their contributions underpin the research that focuses on behavioral and social aspects in the work activities of software engineers, but deep analysis of individual decision-making behavior is underdeveloped.

Software development involves interdependent individuals working together to achieve favorable outcomes, so the decision-making behavior of each individual will influence behaviors of other teammates and the project outcome. Individuals have many chances to make a decision in a development process. For example, individuals may choose a resolution to deal with a conflict. In agile development, each one makes a decision about effort estimation and gives user story points. Individuals may often independently make "work" or "shirk" choices in teamwork. Under these conditions, different individual decision-making behaviors will generate different results, which are pertinent to the success or failure of the project. Therefore, it is imperative to study individual decision-making behavior in SE.

Social cognitive theory emphasizes the bi-directional interactions between three elements: individuals, environment, and behavior. Overt behavior is influenced by these intrinsic and extrinsic factors [4]. This theory provides two directions for a study of individual decision-making behavior: individual and environmental. About the former, there is no doubt that characteristics of individuals exert a strong influence on individual behavior. Personality is regarded as an important internal property. There is a substantial body of research that has sought to explore the effect of individual personality in SE. About the environmental factors, some literature in SE also studies this. Acuña and Juristo [1] argue three environments (organizational, cultural, and technological) were important for managing both the activities and the members of a software project team. Xu and Ramesh [52] gave four aspects of the environment from the perspective of software process tailoring. Hossein and Aybuke [26] presented the environmental factors influencing IT personnel intentions to leave. Clarke and Connor [15] discussed the situational factors that affected the software development process. The description of environmental factors in those studies is either macroscopic, or not oriented to individual decision-making behaviors. There is still a lack of a systematic and deep analysis on the influencing environmental factors of individual decision-making behavior. Various environmental factors, such as task complexity and team cohesion, also exert great influence on individual behavior. So, in order to achieve a desired quality of understanding and prediction of individual behavior, a detailed investigation of the influencing environmental factors of individual decision-making behavior is necessary.

From the perspective of decision theory, it is also necessary and worthwhile to determine which environmental factors influence individual decision-making behavior. In decision sciences domain, the assumptions about individuals have evolved from complete rationality, to bounded rationality, and then to ecological rationality. Initially, the individual is thought to behave as a completely rational person to seek utility maximization during decision making. Then the individual is regarded as operating under bounded rationality due to cognitive limitations, and s/he pursues a satisfactory, but not optimal, solution. Ecological rationality is proposed based on the adaptation theory. Some scholars argue that the decision-making process is influenced by the environment, and individuals tend to have an adaptive characteristic [41]. So, the decision-making behavior of an individual is self-adaptive, resulting from the interaction between the individual and the individual's environment. This theory also supports the importance of studying environmental factors in order to understand individual decision-making behavior.

Therefore, this paper aims to discover which environmental factors will affect individual decision-making behavior in SE. We conducted a SLR to explore the issue. We analyzed the literature that was selected from our intensive search, and identified, summarized, and classified the related factors in the studies. Our analysis presented a comprehensive reference list and taxonomy of the environmental factors affecting individual decision-making behavior in SE.

This article is organized as follows. Section 2 describes the method of SLR used in this paper. The results of SLR are presented in section 3. Finally, we present the discussion on our results, and conclude the paper.

## 2. RESEARCH METHOD
Researchers have often used SLR to answer their research questions [26]. The SLR is a methodical way to identify, evaluate, and interpret the available studies conducted on a topic, research question, or a phenomenon of interest. According to the guideline in [34], Figure 1 gives the detailed steps used in this study.

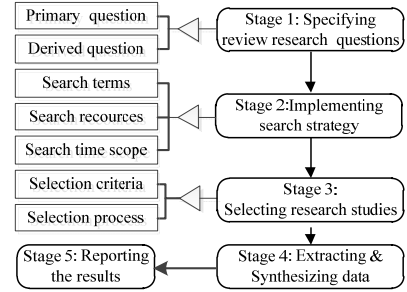

**Figure 1. The detailed steps of SLR.**

### 2.1 Specifying the Review Research Questions
Research questions provide the guidance for the review. Building from the aim of this study, the primary research question is "which environmental factors influence individual decision-making behavior in SE projects?" We can imagine the answers are various and diverse. From the viewpoint of software project management, it is important to classify these factors so as to understand, analyze, and manage them effectively. Therefore, we wanted to know whether there was any classification method of environmental factors in the existing literature. If so, can this method be adopted or improved for this study? If not, how will we classify those factors? Based on the consideration, one derived question is presented "what is the classification method related to environmental factors in SE projects?" The primary and derived questions together drive this SLR.

### 2.2 Search Strategy
The search strategy mainly included three aspects: search terms, search resources, and search time scope.

The search keywords largely determined the quality of search results. Based on the primary research question, a combination of "software engineering," "individual," "decision-making behavior," and "environmental factor" was expected. However, too many strings may bring too small coverage. When using this search expression, we noticed the number of search results were far fewer than we expected: only five records were available. Moreover, most of the results had little relation to our topic. Therefore, we carefully considered each search term in order to reduce the bias and retrieve as many papers as possible. First, we thought "decision-making behavior" could not be regarded as a keyword in spite of its importance in our topic. As mentioned before, developers need to make a decision under many situations, such as cost estimation and development model selection. Literature related to these topics should also be examined to check whether there is mention of any influencing factor. However, there was not an obvious word "decision-making" or "behavior" in the literature generally. So those literature would not be located if "decision-making behavior" was in the search terms. In fact, the words "decision-making" and "behavior" rarely appear in SE field. But they do appear in management science field. Applying decision theory into SE is the aim of this paper. So, the term "decision-making behavior" was excluded from the keywords. Second, individuals in SE are usually called software engineers or developers. Therefore, the two keywords of "software engineering" and "individual" are combined and turned into one phrase of "software engineer/developer." Third, given the synonyms, we added four synonyms of "environmental": situational, external, contextual, and surrounding. Fourth, because motivation/de-motivation factors, which have been discussed recently [11; 31], are closely related to software engineers' decisions, and some motivation/de-motivation factors come from the external environment, "motivation/de-motivation factor" was also added to the search terms. Then we obtained the following search expression by using the operator AND or OR:

("software engineer" OR "software developer") AND ("environmental factor" OR "situational factor" OR "external factor" OR "contextual factor" OR "surrounding factor" OR "motivation factor" OR "de-motivation factor").

In order to perform a broad search, instead of limiting the search sources we used a comprehensive search engine that can search all the databases to which Beihang University in China is subscribed. These databases include ACM, IEEE, ScienceDirect, and so on. The comprehensive search engine can easily travel through all the databases, but it can also produce many irrelevant results. In order to enhance the pertinence of search results, the research results with obviously unrelated subject types were excluded. Because our topic is an overlapping field between SE and management science, we paid attention to the search results in four subject types: technology, social sciences, sciences, and psychology. In addition, in order to reflect a snapshot of the current state of related research in SE, the time period of works for our search was limited to January 2000 to December 2014.

## 2.3 Select Research Studies

Primary studies were selected according to the following selection criteria. We included studies that: 1) directly give some answers to the research questions. 2) are available for full-text reading by the online library service of Beihang University. 3) relate to a decision process or environmental factors in some aspects of SE. 4) relate to individuals of software providers, not users. We excluded studies that were: 1) in languages other than English, 2) duplicated or repeated studies, 3) unrelated Subjects, and 4) in the form of editorial notes, prefaces, or article summaries.

The selection process included three steps and involved two roles (see Figure 2). Firstly, a junior researcher performed the search according to the search strategy, and applied the selection criteria on the title of each result to exclude many studies that were clearly irrelevant. Secondly, the junior researcher further excluded some papers after carefully reading the abstracts. In order to improve the quality of search results, a senior researcher randomly chose one fourth of the studies that had been discarded or reserved by the junior researcher to review and adjust the results so as to get the final results of this step. Finally, the junior researcher selected relevant papers according to a full text reading. Just like the second step, the senior researcher also reviewed and adjusted. But the difference is that the rate of random selection was increased to one third because full text reading needs a more rigorous audit for data extraction. After three rounds of filtering the final search results, 40 primary studies were found.

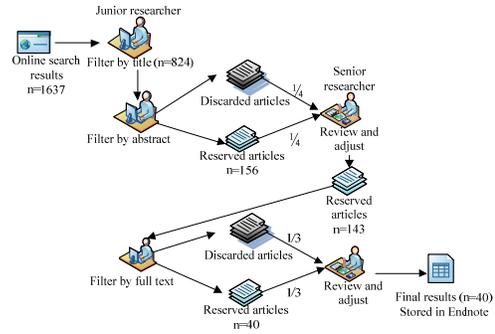

**Figure 2. The selection process.**

## 2.4 Data Extraction and Synthesis

According to the selection sequence, each paper was stored in a literature management software, Endnote X7, and assigned an ID. Because it is easy to get basic information – such as publication year, authors, and type of study – by this software, we focused on extracting data which could directly answer the research questions. Data extraction is based on full text reading, so full text filtering during study selection and data extraction were carried out simultaneously. To carry out this phase more efficiently, three data extraction forms were designed and implemented in MS Excel. In the first worksheet, each paper was marked by Y/N to indicate if a classification of factors existed in it. For a paper with a classification, the ID, the categories, original factors and its total number, factors adopted in this paper, and its number were recorded in the second worksheet. For those papers without a classification, the ID and the factors extracted from each paper were recorded in the third worksheet. Data synthesis included two steps. Firstly, we analyzed all the categories in the second worksheet, merged some categories with the same meaning, and put forward a new taxonomy. Then, we combined the factors in the second and third worksheets into one form. We dealt with this step very carefully. Only these factors with obviously the same meaning in the original papers were combined into one factor. If any small difference between two factors existed, we kept them alone, even though they almost seemed the same. And each factor was associated with a category mentioned in the first step. Then the source, category, and frequency of each factor in all of the papers were recorded, which are reported in section 3.

## 3. RESULTS

The 40 primary studies gained by the SLR and the citation from the reference list are tabulated in Table 1. This representation of primary studies and references is adapted from [32]. From here on, each primary study is referred to using the ID in Table 1.

**Table 1. Studies and references**

| ID | Cit. | ID | Cit. | ID | Cit. | ID | Cit. |
|----|------|----|------|----|------|----|------|
| S1 | [19] | S11 | [25] | S21 | [11] | S31 | [40] |
| S2 | [16] | S12 | [30] | S22 | [55] | S32 | [38] |
| S3 | [46] | S13 | [39] | S23 | [17] | S33 | [43] |
| S4 | [7]  | S14 | [28] | S24 | [35] | S34 | [12] |
| S5 | [24] | S15 | [21] | S25 | [3]  | S35 | [54] |
| S6 | [20] | S16 | [22] | S26 | [8]  | S36 | [23] |
| S7 | [51] | S17 | [50] | S27 | [10] | S37 | [49] |

| S8 | [15] | S18 | [44] | S28 | [9] | S38 | [48] |
| S9 | [53] | S19 | [29] | S29 | [6] | S39 | [47] |
| S10 | [45] | S20 | [31] | S30 | [26] | S40 | [52] |

### 3.1 Overview of the Studies

Figure 3 presents the temporal distribution and type of 40 studies.

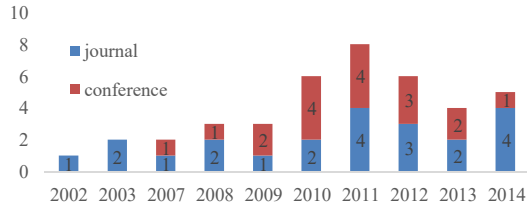

**Figure 3. Temporal distribution and type of primary studies.**

Out of the 40 primary studies, 29 (72.5%) have been published in the last five years. This indicates a growing trend in the importance placed on social aspects research from related disciplines in SE in recent years. Figure 3 also shows that journal publications occupy 55% (22/40) of all primary studies and conference proceedings 45% (18/40). Among those from periodicals, 15 (68%) come from four important journals that are among the top six leading SE journals according to [5]. Among those from conference proceedings, one (5.6%) is from the International Conference on Software Engineering, and three (16.7%) from the workshop on Cooperative and Human Aspects of SE, which is the top workshop about human factors in SE. In addition, some other conferences closely related to SE are also sources, as shown in Table 2.

**Table 2. Some sources of studies**

| Type | Name | No. |
|---|---|---|
| Journal | IEEE Software | 2 |
| Journal | Journal of Systems and Software (JSS) | 6 |
| Journal | Information and Software Technology (IST) | 6 |
| Journal | ACM Transactions on Software Engineering and Methodology (TOSEM) | 1 |
| Conference | International Conference on Software Engineering (ICSE) | 1 |
| Conference | ICSE Workshop on Cooperative and Human Aspects of Software Engineering (CHASE) | 3 |
| Conference | International Conference on Software Maintenance (ICSM) | 1 |
| Conference | Empirical Software Engineering and Measurement (ESEM) | 2 |
| Conference | Evaluation and Assessment in Software Engineering (EASE) | 2 |
| Conference | ACM Special Interest Group on Management Information System - Computer Personnel Research (SIGMIS CPR) | 1 |

The distribution of 40 articles among the database sources is shown in Figure 4. The total number is actually more than 40, because some papers appear in multiple databases.

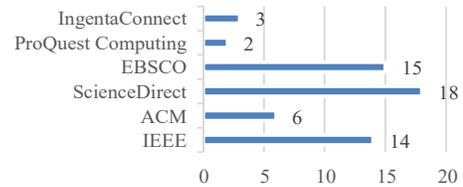

**Figure.4 Database sources of studies.**

### 3.2 Results of Research Questions

We went through each paper and extracted any classification provided in the data source to answer the derived question. Eight out of 40 primary studies provided some classifications and associated factors. However, these categories and factors were not directly oriented to our theme, so not all of the categories and factors were suitable for our paper. On the other hand, although some categories coming from various papers seemed different, they actually expressed the same or similar meaning. Through carefully checking, we selected and incorporated some categories from the eight papers, then attained a new fit taxonomy that includes eight categories. And some factors from the eight papers were selected to answer the primary research question.

Table 3 shows some information about the eight studies and the taxonomy. The number of categories, original factors, and adopted factors of each paper are given in the first four rows. In the following part, the relationship between the new taxonomy and eight studies is shown. The first column shows the new eight categories and their ID. For each category, the symbol "Δ" represents which article has provided the category.

A description of each category is presented here to help us fully understand the taxonomy. C1 represents the characteristics of a software project and the special decision task in the project. C2 represents the knowledge or technology needed for a decision. C3 represents the power of the software engineer to make a decision. C4 represents the balance between work and personal life that will be considered by software engineers while making a decision. C5 represents whether the decision is related to the software engineer's long-term career development, new technologies or knowledge development, and financial reward. C6 represents characteristics of the managerial personnel or methods in the software project. C7 represents the profile of the organization and personnel outside of the project team. C8 represents the profile of the team itself and teammates.

**Table 3. The taxonomy and eight data sources**

| Data sources | S1 | S2 | S5 | S7 | S8 | S9 | S15 | S40 |
|---|---|---|---|---|---|---|---|---|
| No. of categories | 3 | 5 | 5 | 3 | 8 | 2 | 4 | 4 |
| No. of factors | 58 | 20 | 43 | 21 | 44 | 6 | 28 | 20 |
| No. of used factors | 49 | 14 | 28 | 20 | 11 | 4 | 26 | 11 |
| **Classification** | | | | | | | | |
| C1:Task characteristics | Δ | | Δ | Δ | Δ | | Δ | Δ |
| C2:Competence | | Δ | | | | | | |
| C3:Power | | Δ | | | | | | |
| C4:Balance between work & life | | Δ | | | | | | |
| C5:Career | | Δ | | | | | | |
| C6:Managerial Characteristics | | | Δ | Δ | Δ | | | |
| C7: Organization Characteristics | Δ | | Δ | Δ | Δ | Δ | Δ | Δ |
| C8:Team characteristics | Δ | | Δ | | Δ | Δ | Δ | Δ |

Date units representing environmental factors were extracted from each paper. These factors were examined carefully in order to reduce data redundancy. If some factors from different sources obviously have the same meaning, they were replaced by the same one among these factors. Meanwhile, the factor's source was updated to reflect which papers included the factor. And frequency to show the number of occurrences of each factor was calculated. Finally, 237 factors from the 40 studies were identified. And, each factor was assigned to a category according to the meanings of factors and categories mentioned above. Because of space limitations, in this paper only some major factors whose frequencies are bigger than 4 are listed in Table 4. In the first column, the number in parentheses shows the total number of environmental factors belonging to each category. The other columns give the name, sources and frequency of each factor. In addition, the top eight environmental factors, whose frequencies are not less than 10, were highlighted in bold and their orders are given after their names in Table 4.

**Table 4 The dominant environmental factors of each category**

| | Factor | Source | # |
|---|---|---|---|
| C1(58) | **Task identity (2nd)** | S1, S2, S4, S5, S11, S14, S15, S16, S18, S19, S21, S26, S30, S36 | 14 |
| | **Task significance (4th)** | S1, S2, S4, S5, S6, S11, S14, S18, S19, S21, S26, S30 | 12 |
| | Workload | S1, S5, S6, S28, S30, S35 | 6 |
| | Stress/pressure | S4, S11, S18, S21, S30 | 5 |
| | Sufficient resources | S4, S11, S18, S19, S21 | 5 |
| | Task variety | S1, S4, S5, S6, S30 | 5 |
| C2 (20) | Technical competence | S1, S6, S15, S21, S23, S25 | 6 |
| | Creativity | S16, S21, S28, S36, S37 | 5 |
| | Development practice | S11, S16, S19, S21, S36 | 5 |
| C3 (10) | **Autonomy (1st)** | S1, S2, S4, S5, S6, S11, S15, S16, S17, S18, S19, S21, S24, S25, S26, S30, S36, S37, S38 | 19 |
| | **Empowerment (5th)** | S2, S4, S11, S16, S17, S18, S19, S21, S26, S36, S39 | 11 |
| C4 (5) | Work / life balance | S2, S4, S11, S18, S19, S21 | 6 |
| C5 (30) | **Rewards & financial incentives (5th)** | S2, S4, S11, S16, S18, S19, S21, S30, S33, S36, S37 | 11 |
| | Change | S1, S2, S5, S6, S11, S16, S19, S21, S32 | 9 |
| | Career path | S1, S2, S4, S11, S18, S19, S21 | 7 |
| | Benefit | S11, S16, S19, S21, S36, S37 | 6 |
| | Challenge | S11, S19, S21, S25, S38 | 5 |
| | Promotion opportunity | S4, S18, S25, S33, S35 | 5 |
| | Reward system | S4, S6, S11, S18, S21 | 5 |
| C6 (28) | **Recognition(6th)** | S5, S6, S11, S15, S18, S19, S21, S26, S33, S37 | 10 |
| | Good management | S8, S11, S16, S17, S18, S19, S21, S24, S36 | 9 |
| | Commitment | S1, S5, S6, S8, S32 | 5 |
| C7 (52) | Work environment | S1, S4, S7, S11, S18, S19, S21, S27, S32 | 9 |
| | Employee participation | S1, S5, S6, S11, S15, S18, S19, S21 | 8 |
| | Equity | S4, S5, S11, S16, S18, S19, S21, S36 | 8 |
| | Job satisfaction | S17, S19, S26, S30, S31, S33, S38 | 7 |
| | Job security | S4, S11, S14, S18, S19, S21, S37 | 7 |
| | Culture | S7, S8, S18, S21, S23 | 5 |
| | Organizational commitment | S11, S30, S31, S38, S40 | 5 |
| | Risk | S4, S11, S21, S29, S34 | 5 |
| | Sense of belonging | S4, S11, S18, S19, S21 | 5 |
| | Working in successful company | S11, S18, S19, S21, S36 | 5 |
| C8 (34) | **Communication (3rd)** | S1, S4, S6, S10, S11, S13, S17, S18, S21, S26, S28, S32, S39 | 13 |
| | **Feedback(6th)** | S11, S14, S16, S17, S18, S19, S21, S26, S28, S36 | 10 |
| | Team working | S2, S4, S7, S10, S11, S19, S21, S30 | 8 |
| | Appropriate working conditions | S4, S11, S18, S19, S21, S33 | 6 |
| | Trust & respect | S4, S11, S18, S19, S20, S21 | 6 |

From the viewpoint of the total number of factors in each category, C1 has the most factors, and C7 is in the second place. C5, C6, and C8 have almost the same number of factors. The factors belonging to C4 are the lowest. However, from the viewpoint of the top eight high frequency factors, two factors come from C3, although the amount of factors of the C3 is the less. And the top eight factors all does not belong to C7, but the total number of factors in C7 is high. Obviously, the importance of each category does not depend on the total number of its factors. Moreover, task identity and significance are far more important in influencing individual decision-making behavior than the other factors of C1. This is in line with the fact that people tend to be more careful when making a decision on important matters. Additionally, the number of factors of C4 is least and its influence is also weak.

## 4. DISCUSSIONS AND CONCLUSIONS

In this paper, we conducted a SLR to answer our research questions. From the 40 articles, a taxonomy including eight categories is provided. For each category, some major environmental factors influencing individual decision-making behavior in software projects were given.

Because individual decision behavior is adaptive, it is necessary to study which environmental factors influence the individual decision behavior in software projects so as to help managers perceive, understand, and even guide individual behavior. As far as we know, this is the first attempt to review the literature about this topic in a systematic way. Our findings present a general comprehensive reference framework for this field. From the viewpoint of the quantity of factors, the factors belonging to C1 and C7 account for 46% (110/237) of the total number. This indicates that the task itself and the organization will affect the individual decision-making behavior from more aspects than those of other categories. This is also in accord with the fact that the software itself and its external environment are very complex. In addition, the influences coming from a software team are various and must not be overlooked (14%), because the individual in a software project always exists in a team and interacts with other team members. For another, when considering the number of occurrences of each factor, the majority of primary studies support autonomy (19/47.5%). This indicates that full authorization is very necessary. In addition, reward, feedback, and recognition are also given importance by most studies. This is in accordance with the theory of needs. If a person makes a decision, s/he will hope that her or his decision will bring her or him economic or spiritual benefits.

The variety and diversity of factors in our reference list and the taxonomy serve as reminders of the level of complexity involved in software project management. In addition to providing a useful reference listing for the researcher for further study of the individual aspects in SE, our results also provide support for practitioners who are challenged with managing a software team.

Of course, this paper has several limitations. Common limitations are about the possible biases introduced in the selection process and inaccuracies of the data extraction in SLR. We tried to avoid them by using a multistage selection process and an audit method completed by two researchers. Additionally, because the environmental factors come from 40 pieces of literature, and the new taxonomy only comes from eight pieces of literature, the taxonomy may not cover all the factors. The relationship between each factor and its category is determined by a subjective understanding, thus inaccuracies of grouping factors may exist.

The limitations discussed above offer clear paths to further research. It would be useful to analyze and categorize environmental factors influencing individual decision-making behavior by the sophisticated semantic analysis technique, and then get the taxonomy. This is what we are planning to do in the future. In addition, it is worthwhile to identify the relationship between the environmental factors and software development process so as to give the project manager guidance to understand and control the influence of different environmental factors during the different software development stages. Moreover, as mentioned before, the intrinsic, personal factors and extrinsic environmental factors are interactive during the decision process. If we know the external and internal factors, describing the interaction between them is also a research direction.

## 5. ACKNOWLEDGMENTS

This research was supported by the China Scholarship Council (CSC).